\documentclass[journal]{IEEEtran}
\usepackage{graphicx}
\usepackage{url}
\begin{document}

\title{The Topological Processor for the future
ATLAS Level-1 Trigger: from design to
commissioning}
%
%

\author{E. Simioni, S. Artz, B. Bauss, V. B\"{u}scher, A. Kaluza, R. Caputo, R. Degele, K. Jakobi, C. Kahra\\ 
A. Reiss, J. Sch\"{a}ffer, U. Sch\"{a}fer, M. Simon, S. Tapprogge, A. Vogel, M. Zinser\\~\IEEEmembership{Institut f\"{u}r Physik\\
                Johannes Gutenberg Universit\"{a}t\\
                Mainz, Germany}}

\maketitle
\thispagestyle{empty}

\begin{abstract}
The ATLAS detector at the Large Hadron Collider (LHC) is designed to measure
decay properties of high energetic particles produced in the proton-proton collisions.
During its first run, the LHC collided proton bunches at a frequency of 20 MHz, 
and therefore the detector required a Trigger system to efficiently select 
events down to a manageable event storage rate of about 400 Hz. By 2015 the LHC instantaneous
luminosity will be increased up to \textbf{$3\times$$10^{34}cm^{-2}s^{-1}$}: 
this represents an unprecedented
challenge faced by the ATLAS Trigger system. To cope with the higher event rate
and efficiently select relevant events from a physics point of view, a new element will
be included in the Level-1 Trigger scheme after 2015: the Topological Processor
(L1Topo).\\

The L1Topo system, currently developed at CERN, will consist initially of an ATCA
crate and two L1Topo modules. A high density opto-electroconverter (AVAGO miniPOD) drives up to
1.6 Tb/s of data from the calorimeter and muon detectors into two high-end FPGA (Virtex7-690), 
to be processed in about 200 ns. The design has
been optimized to guarantee excellent signal integrity of the high-speed links and low
latency data transmission on the Real Time Data Path (RTDP). The L1Topo receives
data in a standalone protocol from the calorimeters and muon detectors to be
processed into several VHDL topological algorithms. Those algorithms perform
geometrical cuts, correlations and calculate complex observables such as the invariant mass.
The output of such topological cuts is sent to the Central Trigger Processor.
This talk focuses on the relevant high-density design characteristic of L1Topo, which
allows several hundreds optical links to processed (up to 13~Gb/s each) using ordinary
PCB material. Relevant test results performed on the L1Topo prototypes to
characterize the high-speed links latency (eye diagram, bit error rate, margin analysis)
and the logic resource utilization of the algorithms are discussed. 
\end{abstract}


\section{Introduction}
\begin{figure}
\includegraphics[width=88mm]{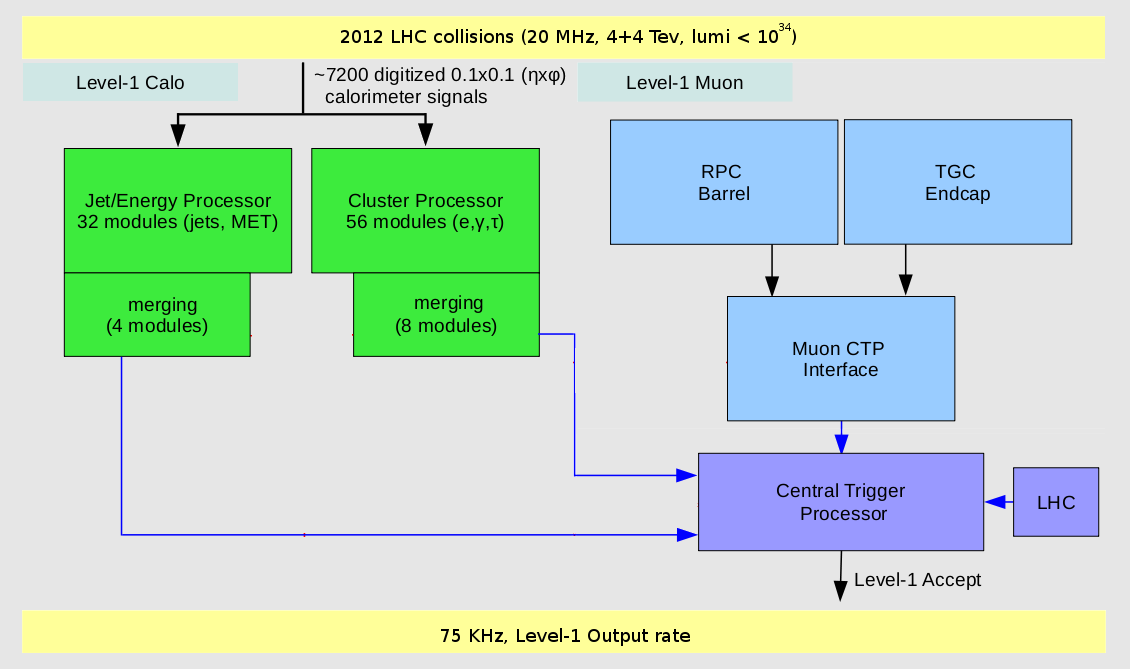}
  \caption{The ATLAS Level-1 trigger systems from the
2010-2013 (Run 1). 
Trigger information goes directly from the detector into the
electronics cavern located in the ATLAS underground area. 
L1Calo: the signals go into
pre-processors and the digitized information is then fed into the Jet/Energy
and Cluster processors. 
L1Muon: signal from RPC and TGC detectors are processed into the MUCTPI processors to identify
muon clusters.
The results are then merged and sent to the Central
Trigger Processor which sends the Level-1 Accept of the event.
}
\label{l1topo_l1}
\end{figure}

\IEEEPARstart{A}{tlas} 
is one of the multi-purpose experiments at the Large Hadron Collider (LHC) at the European
Center of Nuclear Research CERN in Switzerland.
LHC collided bunches of opposing protons at a frequency of 40 MHz. 
The ATLAS Trigger system
filter out collision events without physics interest, 
lowering the average output rate to a level to few hundreds hertz.
It is realized in a
multi level trigger. A level-1 uses 
muon and calorimeter signals
to determine ``Regions of Interest'' (RoI) and based on counting clusters of 
jets, $\tau$, electron/$\gamma$ and missing $E_T$ at various energy threshold, reduce the 
event rate to 75 kHz.
A Level-2 uses  
Level-1 candidates and  
look in detailed physics properties to achieve a further reduction in rate 
to 2-3 kHz. Finally a third level (event filter)
use full event information, and decide upon
storage of the event for offline analisys with a final rate of 300-400 Hz.
This paper is related to the Level-1 upgrade,
further description of Level-2 and event filter can be found here~\cite{atla_trig}.
The ATLAS Level-1 Trigger is a fixed latency, 40 MHz, pipe-lined, synchronous system, built
to operate at the LHC design instantaneous luminosity of $\times$$10^{34}cm^{-2}s^{-1}$.
The Level-1 trigger system consists of three sub-systems:
The Level-1 Calo Trigger (L1Calo)~\cite{l1calo}, the Level-1 Muon Trigger (L1Muon)~\cite{muctpi}, and
the Central Trigger Processor (CTP)~\cite{ctp}.
The hardware of the Level-1 Trigger is primarily based on FPGAs and custom ASICs. 
Including cabling, the maximum latency budget of the
Level-1 electronics chain is 2.5 $\mu s$.
A simplified outline of the Level-1 trigger is shown in Fig.\ref{l1topo_l1}.
In the CTP the trigger decision is made, based on thresholds received from L1Calo and
L1Muon, and forwarded to the higher levels of the trigger.
The L1Calo input data comes from 7200 analog ``trigger towers''. The trigger towers are built
with a granularity of 0.1 $\times$ 0.1 in the $\eta-\phi$ 
range, which is covered by the ATLAS electromagnetic and hadronic calorimeters. 
Digitization of the input signals and digital filtering is done on a
mixed-signal Pre-Processor. Then the trigger tower signals are forwarded to two feature processors, 
namely the Cluster Processor (CP) and the Jet/Energy-sum Processor (JEP). The CP’s aim
is to identify electron, photon, tau and hadron candidates with a Transverse Energy $E_{T}$
above a set of programmable thresholds~\cite{garvey}.
Simultaneously the JEP identifies jet candidates and produces global sums of total, missing, and jet-sum 
$E_T$. Inside the CP and the JEP are the Common Merger Modules (CMMs). 
The CMMs count the multiplicities of trigger objects and send the result
to the Central Trigger Processor (CTP). Upon receiving the ``Level-1 Accept'' signal from the CTP,
the coordinates in the $\eta-\phi$ plane for each trigger object (so called Regions of Interest
, RoIs), which were identified by the feature processors at Level-1, are sent through 
Read-Out Drivers (RODs) on to the Level-2 Trigger.
The L1Muon input data comes from 800 k Resistive Plate Chamber (RPC) strips in the barrel
region and Thin Gap Chambers (TGCs) in the endcap regions. Multiplicities for six thresholds at
high and low transverse momentum are measured by coincident hits in the RPC and TGC planes.
The logic for multiplicity counting of the different thresholds is provided by the Muon Central
Trigger Processor Interface (MUCTPI)~\cite{muctpi}.

\section{Potential in a Level-1 trigger with topological capabilities}
\begin{figure}
\includegraphics[width=88mm]{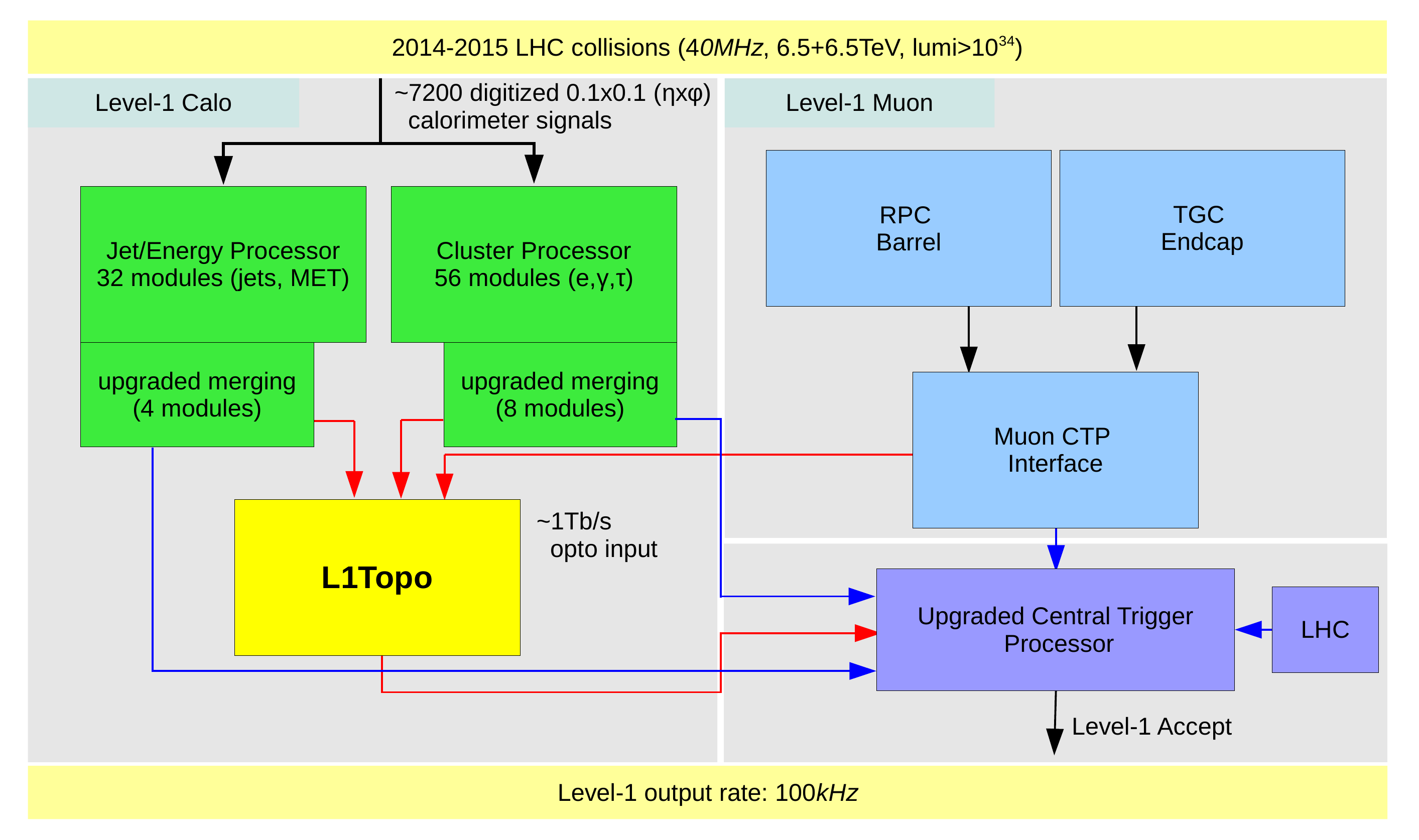}
  \caption{Simplified scheme of the upgraded Level-1 with the 
inclusion of the L1Topo sub-system.}
\label{l1topo_schemex}
\end{figure}

\begin{figure}
\includegraphics[width=88mm]{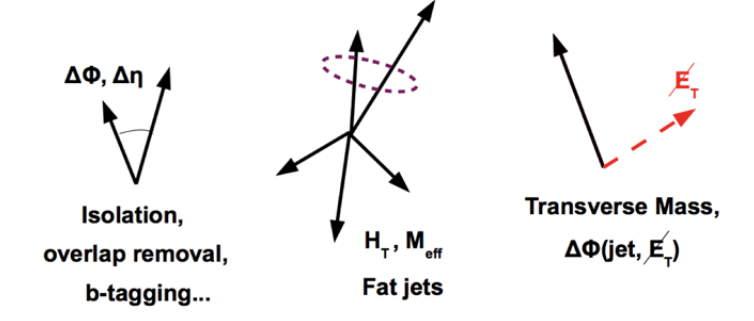}
  \caption{Examples of topologies used for topological trigger decisions. 
Angular distributions are illustrated (left), hardness of the interaction
or sums of energy or momenta (center) and invariant mass can be used to do 
topological cuts.
}
\label{l1topo_topologies}
\end{figure}

From 2015 (Run 2), the increased instantaneous luminosity and collision frequency,
implies that the Level-1 Trigger can't
cope with the higher background rates. To maintain the Level-1 trigger rates at the
current level without unduly raising thresholds or prescaling trigger streams
of physics interest, the selections can't be based just on counting of physics objects.
To achieve additional
background rate reduction at Level-1, the topological information on jet or muon direction in space
can be used. 
For doing so, the current CMMs will
be replaced by new ``CMX''~\cite{cmx} modules capable of receiving
and processing the high-speed backplane data and transmit the real-time
date to a completely new element in the level-1 chain: the Topological
Processor (L1Topo).
The L1Topo will allow a trigger decision to be made using more than just $p_T$
or $E_T$ whose current thresholds will be impossible to maintain in 2015 after the LHC upgrade. 
This is essential 
because interesting physics events have specific 
topologies (see Fig.\ref{l1topo_topologies}).
These decisions can be loosely assigned to three
categories defined as: angular separation, invariant
mass, and hardness of interaction.
To be able to form trigger decisions based on topological information, an entirely new element
in the Level-1 Trigger is required: the Topological Processor (L1Topo, see Fig. \ref{l1topo_schemex}). L1Topo will provide high
optical input bandwidth and powerful state of the art FPGAs, receiving and processing L1Calo and L1Muon
information within 200 ns. 
%
%

\subsection{Topological algorithms}
The L1Topo process real-time event information based on the geometric and 
kinematic relationships between Trigger OBjects (TOBs) (i.e. electrons
photons, muons, jets, and taus), as well as event-level quantities such as 
missing transverse energy, inavariant mass etc.
Algorithms are implemented in VHDL and operated in parallel into FPGAs.
L1Topo input data formats, the number of TOB and corresponding input fibers from 
L1Calo and L1Muon subsystem have been defined and so the data transmission protocol.
In parallel with the hardware development, simulation studies on possible
successfull topology are going on.
Currently the proposed L1Topo algorithms have been implemented in VHDL and
only about 30$\%$ of the resources of the L1Topo system are so far utilized.
More input from physics working group is therefore expected.  

\section{L1Topo system}
In this section the requirements of the L1Topo system and a brief description of the L1Topo
module design is given.
\subsection{L1Topo requirements}
The L1Topo module has to meet the target data rates at
the external interfaces: 6.4~Gb/s on the optical real-time I/O.
Total link bandwidth and logic resource availability are fixed by the
components chosen (largest devices on the market at time of detailed
design work) and cannot be changed. The availability of resources was
checked against the needs of the physics algorithms. 
Given the scalability of the system, more requirements imply more module.
At current, an L1Topo system equipped with two modules will satisfy 
the requirements during Run2.
A non scalable requirement relate to the total latency of L1Topo system.
The ATLAS latency envelope for Run2 
imply a total latency of about 10 LHC bunch crossing (BX). 
Data receiving and transmission on Multi Gigabit Tranceiver (MGT) will require
4 BX. 
Deserialization into the LHC bunch clock domain takes one BX.
The algorithmic latency
is required to be of three BX or lower. All algorithms projected
and implemented so far have been kept to three bunch ticks or be-
low. CRC check sum decoding 
imply an extra 0.25 BX.

\subsection{L1Topo implementation}
\begin{figure}
\includegraphics[width=88mm]{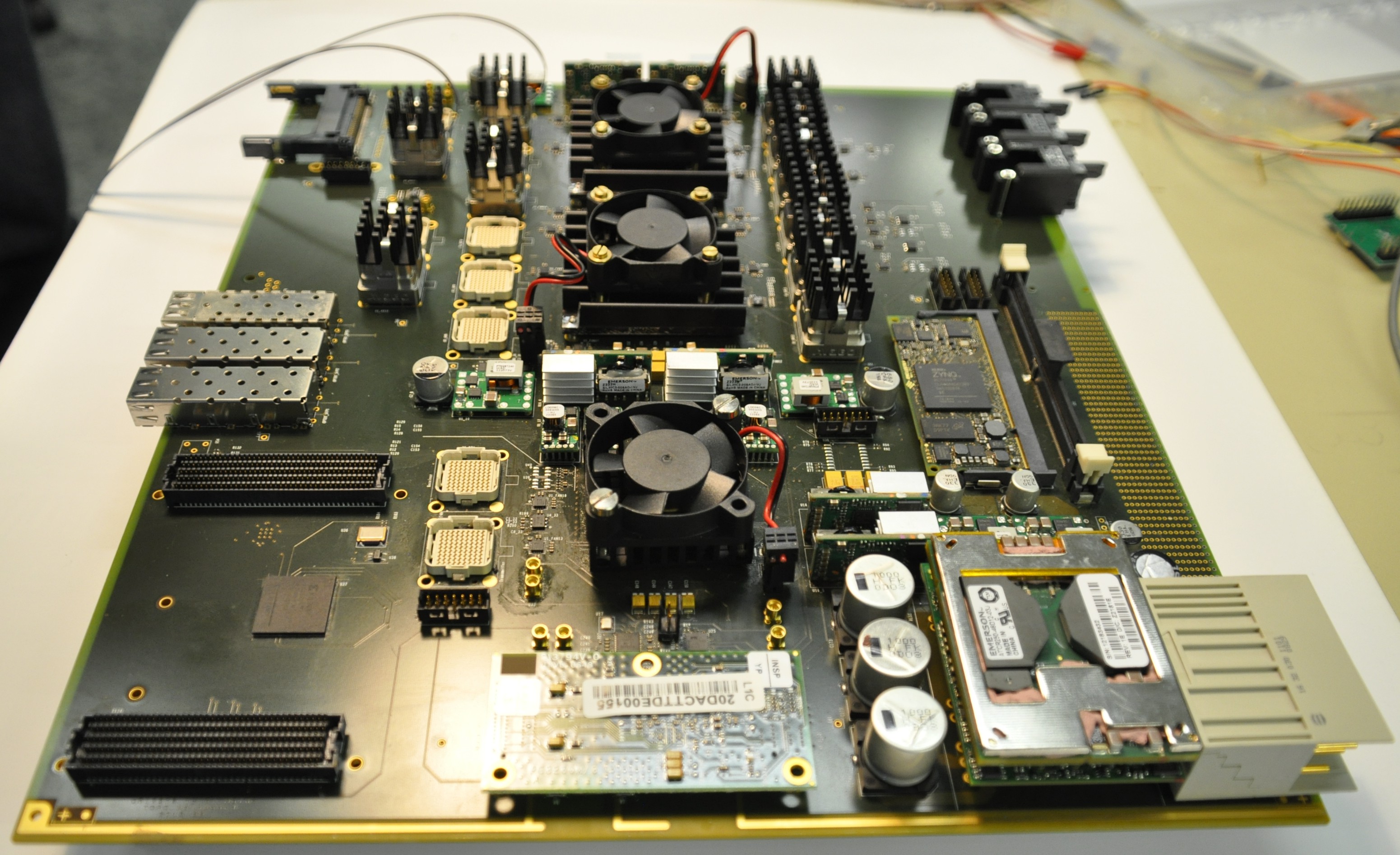}%
  \caption{L1Topo prototype.}
\label{l1topo_prototype}
\end{figure}

The Topological Processor will be a single processor crate equipped 
with two processor modules.
The processor modules will 
be identical copies, with firmware adapted to the specific topologies 
to operate on.\\
Data are transmitted on 
optical fibres. After conversion to electrical representation, data are received 
and processed in FPGAs equipped with on-chip Multi-Gigabit Transceivers (MGT). 
In Run2 results are sent to the Central Trigger processor (CTP) electrically from
the front panel to reduce latency.  
The L1Topo module will be designed in AdvancedTCA (ATCA form factor). A picture of a prototype is shown in Fig.\ref{l1topo_prototype}.

\subsubsection{Real-time data path}
ATCA Backplane zone 3 of L1Topo is used for real-time data transmission. 
The input data enter L1Topo optically through the backplane. 
The fibres are fed via four to five blind-mate backplane connectors that can carry 
up to 72 fibres each. 
In the baseline design 48-way connectors will be used. 
The optical signals are converted to electrical signals in 12-fibre receivers. 
For reasons of design density miniPOD receivers will be used. 
The electrical high-speed signals are routed into two FPGAs, where they are 
de-serialized in MGT receivers; the parallel data are presented to the FPGA fabric. 
The two FPGAs operate on their input data independently and in parallel. 
High bandwidth, low latency parallel data paths allow for real-time 
communication between the two processors. The final results are transmitted 
towards the CTP on both optical fibres and electrical cables. 
The electrical signals are routed via an extension mezzanine module.

\paragraph{Data reception}
The optical data arrive on the main board on 12-fibre ribbons. 
Since the backplane connectors support multiples of 12 fibres, 
the optical signals will be routed via ``octopus'' cable sets, splitting 48
fibres into groups of 12. 
The opto-electrical conversion will be performed in Avago miniPOD 12-channel devices. 
The opto receivers exhibit programmable pre-emphasis so as to allow for 
improvement on signal integrity for given track length. 
After just a few centimeters of electrical trace length, the multi-gigabit 
signals are de-serialized in the processor FPGAs. 
They allow for programmable signal equalization on their inputs. 
The exact transmission protocol is defined and include standard 8b/10b encoding 
(envisaged for purpose of run length limitation and DC balance) and a baseline speed of 6.4~Gb/s. 
The processors are supplied with required bias and termination voltages, 
as well as suitable reference clocks. 

\paragraph{Data processing}
Topology data are processed in two XC7VX690T FPGAs. There is no data duplication 
implemented at PCB level. Therefore different schemes can be employed. 
The two processors can communicate via their fabric interface to get 
access to data that cannot be received directly via the multi-gigabit links. 
Though according to the device data sheets higher data rates should be possible, 
a maximum bit rate of 1~Gb/s per differential pair is anticipated for the 
inter-FPGA link. That will limit parallel connectivity to 238~Gb/s of 
aggregate bandwidth. This would correspond to 24$\times$238 bits per BX (5712 bits)
which allows for sharing more than 250 generic trigger objects (TOBs).
This is more than the outputs of all of the sort trees combined.
Since this data path adds approximately one bunch tick of 
latency, it might be more attractive to fan out data electrically or optically 
at the source, so that both processors are supplied with the same data. 
Due to the large amount of logic resources in the chosen FPGAs, 
a signiﬁcant number of algorithms are expected to be run in parallel on real-time data.
The expected output to the CTP 32 bit per processor, indicating whether a specific 
topological algorithm passed or not plus optionally an overflow bit. 
The resulting trigger data are expected to exhibit a rather small volume.  
They will be transmitted to the CTP also optically.
As required by future upgrades post Run2, a single fibre-optical ribbon 
connection per processor FPGA, running through 
the front panel of the module, is provided for this purpose. 

\subsubsection{Clock distribution}
The operation of the real-time data path requires low-jitter clocks throughout the system. 
For synchronous operation, data transmitters will have to be operated with 
clean multiples of the LHC bunch clock. 
The L1Topo module will be designed for 40.0789 MHz operation of the 
real-time data path only.
The MGTs are clocked off multiples of the LHC clock. 
The jitter on the MGT bunch clock path is tightly controlled with help of a PLL device. 
The clock fan-out chips chosen are devices with multiple signal level input compatibility, 
and output levels suitable for their respective use, either LVDS or CML.

\paragraph{Pre-configuration access and module control}
L1Topo is a purely FPGA-based ATCA module. 
All communications channels are controlled by programmable logic devices 
and become functional only after successful device configuration. 
The prospective ATLAS standard (LAPP IPMC / mini-DIMM format) will be mounted on L1Topo. 
An initial step of module initialization is performed by an IPMC device. 
It communicates to the shelf via an I2C port (IPMB) available on all ATCA modules 
in zone 1 (see below). 
On standard ATCA modules, IP connectivity is mandatory for module control. 
This is provided by two 10/100/1000
Ethernet ports on the backplane in zone 2 (redundant base interface)
wired 
via an SGMII Phy device.

\paragraph{FPGA configuration}
The baseline (legacy) FPGA configuration scheme on L1Topo is via a 
CompactFlash card and the Xilinx System ACE chip.
The required board-level connectivity will be available on L1Topo, 
to write the flash cards through a network connection to the FPGAs, 
once they are configured. 
A local SPI memory and SD card are also wired in the design to provide 
alternative configuration methods. Both devices are placed on a mezzanine
to adapt for future changes. 

\paragraph{Monitoring and control}
The default ATCA monitoring and control path is via I2C links in zone 1. 
The backplane I2C port (IPMB) is connected to the IPMC DIMM. 
On L1Topo configured FPGAs can be monitored for die temperature and internal supply voltage.

\section{L1Topo commissioning}
\begin{figure}
\includegraphics[width=88mm]{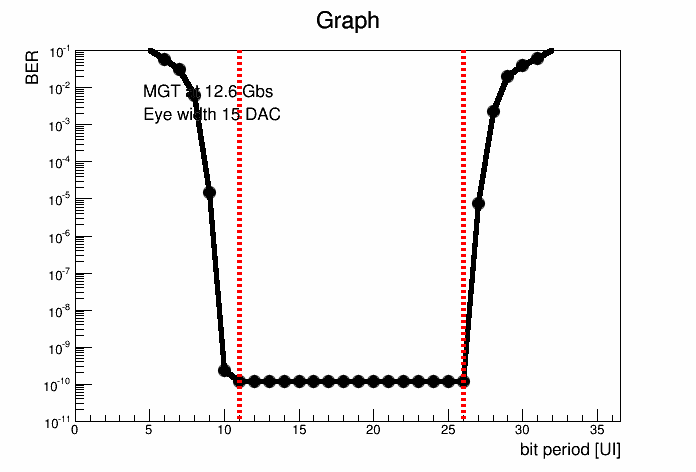}%
\caption{Bathtube profile obtained from MGT loopback on XC7VX690T mounted on the second 
L1Topo prototype. Scan of 1 second per step at 12.8~Gb/s.}
\label{l1topo_berscan}
\end{figure}

The production modules will be subjected to rigorous tests. The test procedure was developed
based on the firmware produce in testing the prototipes. After initial smoke tests the 
modules will be boundary scanned, the high speed links will be IBERT~\cite{ibert} tested for
data eye width and bit error rate. Finally a data transmission test 
with production firmware and final data formats will follow.\\
The latter tests, comprising all high speed interfaces (including readout links) will
take place at CERN. Please note that initial
link tests will for the production modules be performed at 12.8~Gb/s (see Fig.\ref{l1topo_berscan}),
whereas the full module tests can be done at the pre-phase1 target bit
rate of 6.4~Gb/s only. 

\subsection{High Speed Links}
The first prototype is equipped with two XC7VX485 devices. 
These device type supports data rates (in terms of LHC clock multiples) up to 
10.2~Gb/s (56 lines per device). 
Five out of 112 links resulted faulty due to a defect on the vias of the PCB plus a 
rework on one of the devices due to an alignment error during the assembly. 
For the remaining 107 links, the performance have been tested at 10~Gb/s by means 
of eye diagrams and bathtub profiles. Error free data transmission 
was observed and the bathtub profile open above 30\%. 
The upper limit on aggregate data transmission was down to BER $\le$ $\times~10^{-17}$). 
Systematic tests to optimize MGT and miniPOD parameters have been performed.
The second prototype has been assembled and is currently under test 
in the home lab. It mounts one large XC7VX690T-3 embedding higher 
performing transceivers up to 12.8~Gb/s. Along with a suitable speed grade 
of the Avago MiniPOD device attached, data transmission tests at 12.8~Gb/s 
have been performed with realistic fibre lengths and configurations. 
With the new PCB printed with no defected vias, all links are working 
error free. 
Further long term testing is ongoing, and at the time of writing this 
paper bit error rates have been measured down to
($\le$ $10^{-16}$).
Not a single error has been observed.

\subsection{Power}
Power consumption tests performed on the first prototype have shown a power usage
close to the limit of the low voltage power regulators (10 Amp) initially adopted. 
Regulators have been replaced with higher power (20 Amp). 
With the new supplies an improvement on the power ripple well below 10 mV was measured. 
On the second prototype precise measurements on the power consumption were carried out. 
In such measurements, an electrical fanout has been designed in such a way to provide 
electrical input to all 14 Avago miniPOD RX sockets. 
The current drained at different data transmission speed was then monitored. 
Power tests have been performed at the time of writing this paper at  12.8~Gb/s as well,
and they suggest 
a power consumption consistent with the later version of Xilinx power estimator spreadsheet
and well within the supply capabilities.
Such test with high speed input 
is way beyond the TDR baseline. 
%
%

\subsection{Control}
IPbus has been implemented and test for errors in repeatedly
read and write registers. Tests have been perfomed at 400 MHz DDR 
between the control FPGA and one of the processors FPGA (XC7VX690T)
setting an initial limit on the error rate BER $\leq~10^{-15}$.

\subsection{L1Muon to L1Topo data transmission}
\begin{figure}
\includegraphics[width=88mm]{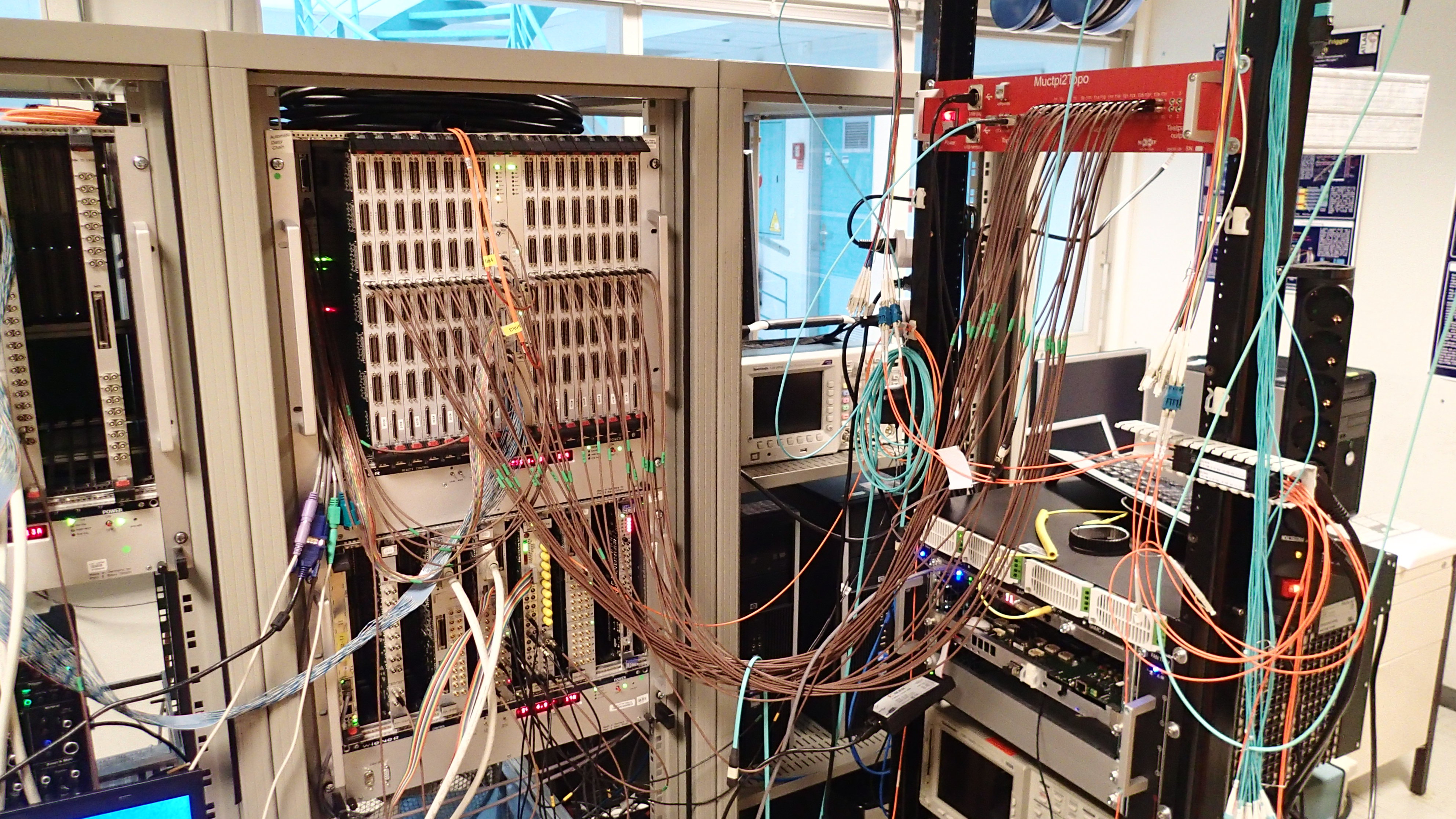}%
\caption{L1Topo under tests with input from the L1Muon system.}
\label{l1topo_powerripple}
\end{figure}

The MUCPTI interface has been installed at CERN and the optical link has been connected with the L1Topo board in a realistic experimental scenario (including optical splitters, see Fig.\ref{l1topo_powerripple}).
Data transmission and signal integrity has been tested by means of statistical Eye 
diagram and bathtub profiles at 6.4~Gb/s. Error free data transmission is observed 
on 4 lines over a period of 24h (BER $\le$ 4$\times~10^{-16}$). 
bathtub profiles have shown an average opening above 30\%.

\subsection{L1Calo to L1Topo data transmission}
Data transmission and signal integrity has been tested by means of statistical Eye
diagram and bathtub profiles at 6.4~Gb/s. Error free data transmission is observed
and bathtub profiles shown an excellent average opening above 55\%.

\subsection{Next steps}
The commissioning phase started few months ago. The tests described above involved 
L1Topo prototypes only. Connectivity to external interface will soon involve the upgraded 
CTP and the existing readout system. The final production module are expected to
exist from September 2014. Tests in the underground electronics racks will begin from the
end of August or before (using prototypes). The deadline for the system commissioning is 
fixed by the end of the year. In this paper we have not aimed to discuss, the software and
firmware effort behind the control, configuration and monitoring of the L1Topo system.

\section{Conclusions}
In this document, an overview of the L1Topo system is presented.
The L1Topo is 
a completely new
element of the ATLAS Level-1 Trigger. 
With L1Topo it will be possible for the first time to apply
topological cuts at Level-1 using detailed information from calorimeters 
and muon sub-detectors. 
With L1Topo it will be possible to apply topology cuts at Level-1 using TOBs
from the calorimeter and muon sub-detectors. 
The L1Topo required additional firmware and hardware upgrades to the existing 
L1Calo and L1Muon systems
which have been here only marginally mentioned or overlooked. 
L1Topo will receive input 
from new CMX modules which are part of
the L1Calo upgrade. It will then use these inputs to make topological based decisions, thus allowing much important
physics to be saved from the alternative of raising the $p_T$ and $E_T$
thresholds. 
Real-time L1Topo output will be sent to the Central Trigger Processor CTP, where the final Level-1
decision is taken. The crucial aspect here is the design of the topology algorithms, including optimization of bandwidth,
FPGA resources and latency. 
The L1Topo prototype is being tested in the L1Calo test lab at CERN 
and has undergone initial integration tests. 
%
The L1Topo production modules are
scheduled for installation on site in 2014.




\end{document}